\documentclass[final,sort&compress,5p,reprint,preprintnumbers]{elsarticle}
\usepackage{graphicx}% Include figure files
\usepackage{dcolumn}% Align table columns on decimal point
\usepackage{bm}% bold math
\usepackage{amsmath}
\usepackage{multirow}

\usepackage[colorlinks=true]{hyperref}
\usepackage{amsmath}
\usepackage{amssymb}
\usepackage{array,booktabs}
\usepackage{graphicx}
\usepackage{subcaption}

\begin{document}
\begin{frontmatter}
%	\title{Kick Amplitude Simulations for Optical Stochastic Cooling using Synchrotron Radiation Workshop}
\title{Wave-Optics Modeling of the Optical-Transport Line for \\
Passive Optical Stochastic Cooling}

\author[niu]{M.B. Andorf}
\author[apc]{V.A. Lebedev}
\author[niu,apc]{P. Piot}
\author[apc]{J. Ruan}
\address[niu]{Department of Physics and Northern Illinois Center for Accelerator \& Detector Development, \\
Northern Illinois University, DeKalb, IL  60115, USA}
\address[apc]{Fermi National Accelerator Laboratory, Batavia, IL  60510, USA}

	
	
\begin{abstract}
 Optical stochastic cooling (OSC) is expected to enable fast cooling of dense particle beams. Transition from microwave to optical frequencies enables an achievement of stochastic cooling rates which are orders of magnitude higher than ones achievable with the classical microwave based stochastic cooling systems. A subsytem critical to the OSC scheme is the focusing optics used to image radiation from the upstream ``pickup" undulator to the downstream ``kicker" undulator. In this paper, we present simulation results using wave-optics calculation carried out with the {\sc Synchrotron Radiation Workshop} (SRW). Our simulations are performed in support to a proof-of-principle experiment planned at the Integrable Optics Test Accelerator (IOTA) at Fermilab. The calculations provide an estimate of the energy kick received by a 100-MeV electron as it propagates in the kicker undulator and interacts with the electromagnetic pulse it radiated at an earlier time while traveling through the pickup undulator.
\end{abstract}
	
	
\end{frontmatter}

\section{Introduction}
\begin{figure}[hhhh!!!!!!]
	\centering
	\includegraphics*[width=0.45\textwidth]{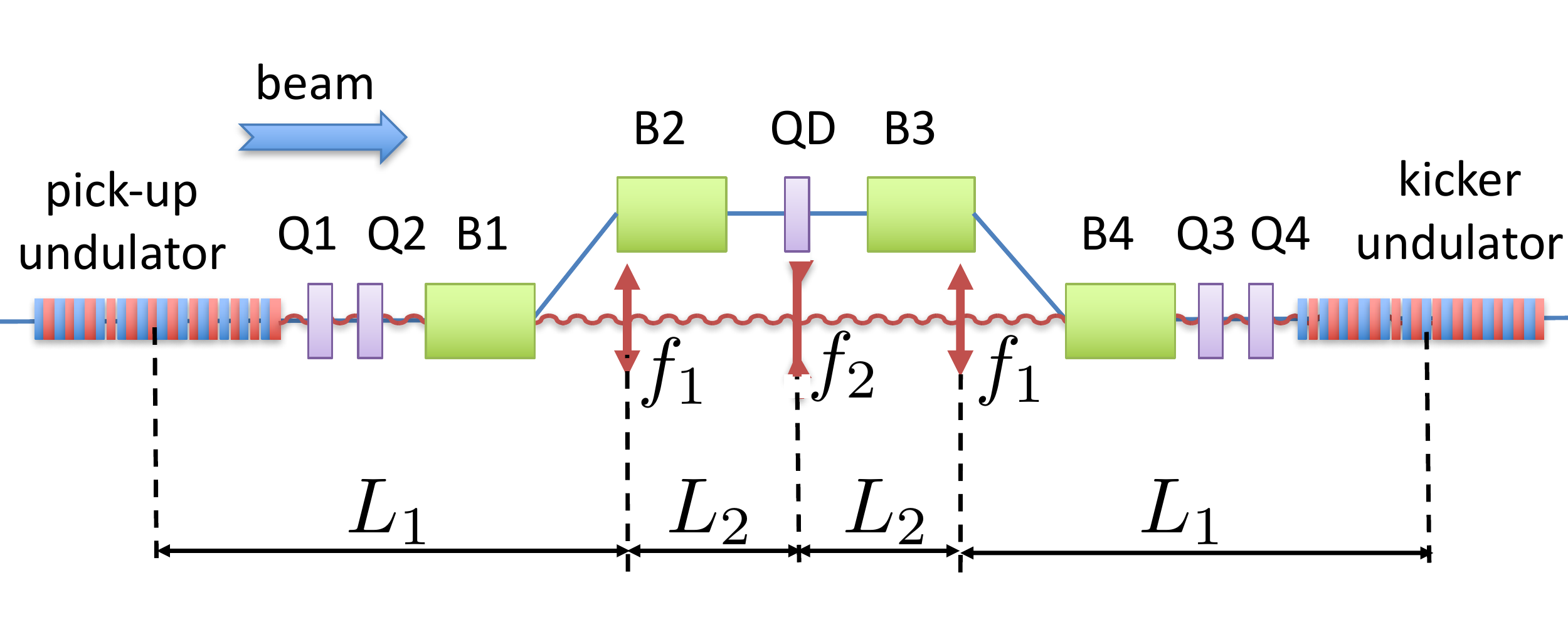}
	\caption{{Overview of the passive-OSC insertion beamline. The labels ``Q$_i$", and ``B$_i$" respectively refer to the quadrupole and dipole magnets and ``$f_i$" represent the optical lenses. The solid blue (resp. ondulatory red) line gives the electron (resp. light) trajectory.}}
	\label{fig0}
\end{figure}

The optical stochastic cooling (OSC) is similar to the microwave-stochastic cooling. It relies on a time dependent signal to carry information on the beam distribution and apply the corresponding cooling force~\cite{OSC_Mikhailichenko,OSC_Zolotorev}; see Figure~\ref{fig0}. In OSC a particle radiates an electromagnetic wave while passing through an undulator magnet [henceforth referred to as the pickup undulator (PU)]. 
The radiation pulse passes through a series of lenses and an optical amplifier and is imaged at the location of a downstream undulator magnet  dubbed as  kicker undulator (KU). The particle beam propagates through a bypass chicane (B$_1$, B$_2$, B$_3$, B$_4$) which provides an energy-dependent path length (i.e. time of flight) as well as a path length variation due betatron oscillations. The chicane also provides the space to house the optical components necessary for the optical-pulse manipulation and amplification. The imaged  PU-radiation field and the particle that radiated it copropagate in the KU resulting in an energy exchange between them. When the time of arrival is properly selected  a corrective energy kick is applied resulting in damping of the particles synchrotron oscillations as the process is repeated over many turns in a circular accelerator. If the KU is located in a dispersive section the corrective kick can also yield transverse cooling in the dispersive plane.  Furthermore if the horizontal and vertical degrees of freedom are coupled outside of the cooling insertion the OSC can provide 6D phase-space particle cooling. 

Although the nominal OSC scheme discussed in most of the literature involves an optical amplifier, the experiment planned in the 100-MeV IOTA electron ring at Fermilab~\cite{IOTA,IOTA2} will not incorporate an optical amplifier in its first phase.  This latter version of OSC is referred to as passive OSC (POSC) and it is considered throughout this paper. The nomination (amplified) OSC scheme will be implemented in a subsequent stage~\cite{andorfAMPLI}. \\

A comprehensive treatment of the OSC can be found in Ref.~\cite{OSC_val} where the kick amplitude is computed semi-analytically by considering a single focusing lens placed between the two undulators separated by a distance much larger than their length. In doing so the depth of field associated with the finite length of the undulators is suppressed. Although theoretically convenient, this focusing scheme is not practical and a three-lens configuration is instead adopted with focal lengths $f_i$ and distances $L_i$ fulfilling~\cite{OSC_val}
\begin{align}
f_1=L_2 &\mbox{~and~} &f_2=-\dfrac{L_2^2}{2(L_1-L_2)} , 
\end{align}
where the parameters are defined in Fig.~\ref{fig0}. The  resulting transfer matrix between the KU and PU defined in  the position-divergence coordinate system $\pmb X =(x,x')$ is $M_{KU\rightarrow PU} =-I$, where $I$ is the $2\times 2$ identity matrix.  The three-lens telescope configuration supports  a longitudinal point-to-point imaging  between the PU and KU while also flipping the transverse coordinate w.r.t. the horizontal kicker axis. Correspondingly the telescope addresses the depth-of-field issue and the results derived for a single lens are directly applicable. The parameters of the optical telescope and undulators (the PU and KU are identical) are listed in Tab.~\ref{tab:TLT}.  Note that both undulators are providing a vertical magnetic field $\pmb B=B\hat{y}$ so that the oscillatory trajectory lies in the $(x,z)$ plane. The undulator radiation wavelength depends on the angle as: $\lambda_r=\frac{\lambda_u}{2\gamma^2}\left[ 1+\frac{K_u^2}{2}+(\gamma\theta)^2\right]$ where the parameters are defined in Tab.~\ref{tab:TLT} and $\theta$ is the observation angle w.r.t the electron direction. Specifically, we defined the on-axis resonant wavelength as $\lambda_0\equiv\lambda_r(\theta=0)$. 
\begin{table}[htb]
\centering
\begin{tabular}{l c l}
			\toprule
			parameter, symbol & value & units \\
			\midrule
			%\toprule
			drift $L_1$     & 143 & cm \\
			focal length $f_1$   & 143 & cm \\
			drift $L_2$         & 32 &cm  \\
			focal length $f_2$       & -4.61 &cm\\
			angular acceptance $\gamma \theta_m$     & 0.8   &        \\
			\midrule
		undulator parameter $K_u$         & 1.038 & \\
		undulator length  $L_u$    & 77.4 & cm           \\
		undulator period,  $\lambda_u$    & 11.057 & cm           \\
		number of periods, $N_u$    & 7 & \\
		on-axis wavelength, $\lambda_0$         & 2.2 & $\mu$m \\
			\midrule
		electron Lorentz factor, $\gamma$    & 195.69 & \\
			\bottomrule
\end{tabular}
\caption{Parameters for the optical telescope and undulators for the proposed POSC experiment at IOTA. \label{tab:TLT} }
\label{param}
\end{table}

\section{Single-lens focusing }
A wave-optics model of single-lens focusing was implemented in the {\sc Synchrotron Radiation Workshop} (SRW) program~\cite{SRW_chubar,SRW_chubar2} to benchmark our numerical implementation with the analytical model obtained for a single lens configuration~\cite{OSC_val}.

\begin{figure}
	\centering
	\includegraphics*[width=0.50\textwidth]{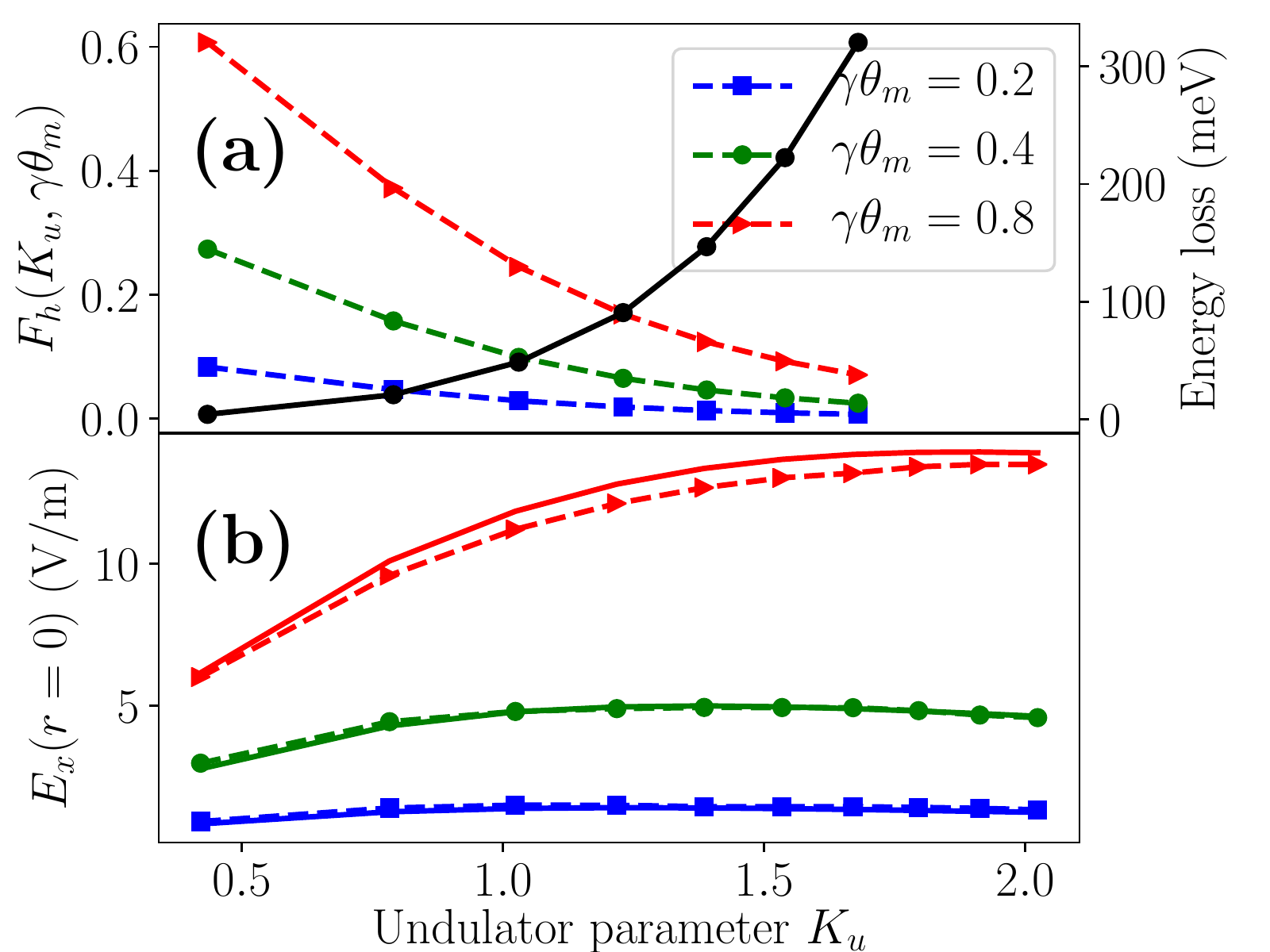}
	\caption{Computed suppression factor $F_h(K_u,\gamma \theta_m)$ (dashed traces, left axis scale) and energy loss (solid traces) of the particle passing through one undulator (right axis scale) as a function of the undulator parameter $K_u$ for different angular acceptances of the lens ($\gamma\theta_m$) (a). Comparison of the electric field at the focus on a single lens analytically computed (solid traces) and simulated with SRW (symbol with dashed traces) for the same cases of angular acceptance (b). }
	\label{fig1}
\end{figure} 

Considering the case of POSC, taking $K_u \ll 1$, and assuming an infinite numerical aperture of the focusing lens, the on axis electric field amplitude imaged in the KU is given by
\begin{align}
E_x(x=y=0)=\frac{4}{3} e K k_u^2\gamma^3, 
\label{small_K_field_amp}
\end{align}
where $k_u\equiv 2\pi/\lambda_u$ ($\lambda_u$ is the undulator period) and $\gamma$ the Lorentz factor. The transverse velocity of the particle is $v_x=\dfrac{K c}{\gamma}\sin(k_uz)$ and the kick amplitude is approximately
%%%
\begin{align}
\Delta {\cal E} =e\int_{0}^{L_u}  \dfrac{E_xK_u}{\gamma}\sin^2{(k_u z)}dz = \dfrac{eE_xK_u L_{u}}{2\gamma}, 
\end{align}
where $L_{u}$ is the undulator length. Combining the latter equation with Eq. \ref{small_K_field_amp} yields
\begin{align}
\Delta  {\cal E}=\frac{2 \pi}{3}(eK\gamma)^2 k_u N_u, 
\label{small_K_kick}
\end{align} 
where $N_u$ is the number of undulator periods. Intuitively Eq. \ref{small_K_kick} is just equal to the total energy loss as the electron travels through one undulator. When $K_u$ is increased (thereby resulting in an increased angular deflection) and the finite angular acceptance of the lens, $\theta_m$, taken into account, the on-axis electric field $E_x(x=y=0)$ in the KU is reduced by a factor $F_h(K_u,\gamma \theta_m)\leq1 $. The expression of $F_h(K_u,\gamma \theta_m)$ is derived in \cite{OSC_val} and its dependence on $K_u$ appears in Fig.~\ref{fig1} for three cases of $\gamma \theta_m$. There is an additional efficiency factor, $F_u(\kappa_u)=J_0(\kappa_u)-J_1(\kappa_u)$, which accounts the effect of the longitudinal oscillation [given by $\frac{K_u^2}{8\gamma^2k_u}\sin(2k_u z)$] of the particle in the KU where $\kappa_u\equiv K_u^2/4(1+K_u^2/2)$. The kick amplitude from Eq. \ref{small_K_kick} is thus reduced by the factor of $F_h(K_u,\theta_m\gamma)\times F_u(\kappa_u)$.\\

The simulation in SRW are performed in the frequency domain: the field frequency components within the PU-radiation bandwidth are propagated and the field amplitude in the time domain inside the KU is computed \cite{light_optics_ipac16}. This is first done for the case of a single focusing lens using $L_u$ and $\lambda_o$ from Tab.~\ref{tab:TLT}, but varying $N_u$ and other parameters appropriately. For this benchmarking simulation, the distance between the PU and  KU centers is taken to be $L_t=19.5$~m (i.e. $L_t\gg L_u$) in order to suppress the depth-of-field effect and the focal length of the lens is $f=L_t/2$. The simulated value for $E_x(K,\gamma \theta_m)$  are found to be in excellent agreement (relative difference below 5\%) as shown in Fig.~\ref{fig1}.

\section{Imaging with a three-lens telescope}

We now focus on the imaging scheme  proposed for the POSC experiment at IOTA with parameters summarized in Tab.~\ref{tab:TLT}. The point-to-point imaging of the KU radiation in the PU is accomplished with a three-lens telescope. First, the field amplitude at the KU longitudinal center is compared with the expected value from theory: using Eq.~\ref{small_K_field_amp} and $F_h(1.038,0.8)=0.25$ yields $E_x=11.8$~V/m while SRW gives 10.9 V/m corresponding to  a relative discrepancy $<7\%$. The kick amplitude using Eq.~\ref{small_K_kick} and $F_u(0.18)=0.91$ yields $\Delta {\cal E}=22$ meV while directly computing the kick in the same way with SRW yields a value of $20.1$ meV. Therefore the agreement between theoretical predictions and numerical simulations is reasonable as was already observed in the previous Section.  

It should be noted that with SRW the longitudinal and transverse dependence of the electric field neglected in theory can also be accounted. The latter of which is from the effective aperture of the outer lenses being less at the edges of the undulator than they are at the center. To find the kick value from SRW, the time-domain field was computed along the kicker every 3.2 mm. The average forward velocity of the particle is $\langle v_z\rangle \equiv \bar{\beta}c=c\beta(1-K_u^2/4\gamma^2)$ where $c$ denotes the velocity of light. Therefore as the particle advances through the kicker it falls back relative to the radiation packet by an amount
\begin{align}
\delta_t = \frac{z_l(1-\bar{\beta})+\frac{K_u^2}{8\gamma^2k_u}\sin{(2k_u z)}}{c}, 
\end{align}
with $z_l$ the location of radiation packet in the KU referenced to its entrance. The latter equation, which also accounts the electron's longitudinal oscillatory motion, is used to compute the electric field $E_x(x,z)$ experienced by the electron  as it propagates through the KU. The  change in energy is then obtained via the numerical integration of:
\begin{align}
\Delta {\cal E} = \int_{z=0}^{z=L_u} v_x E_x(x,z) dz.
\end{align}
\begin{figure}[h]
	\centering
	\includegraphics*[width=0.495\textwidth]{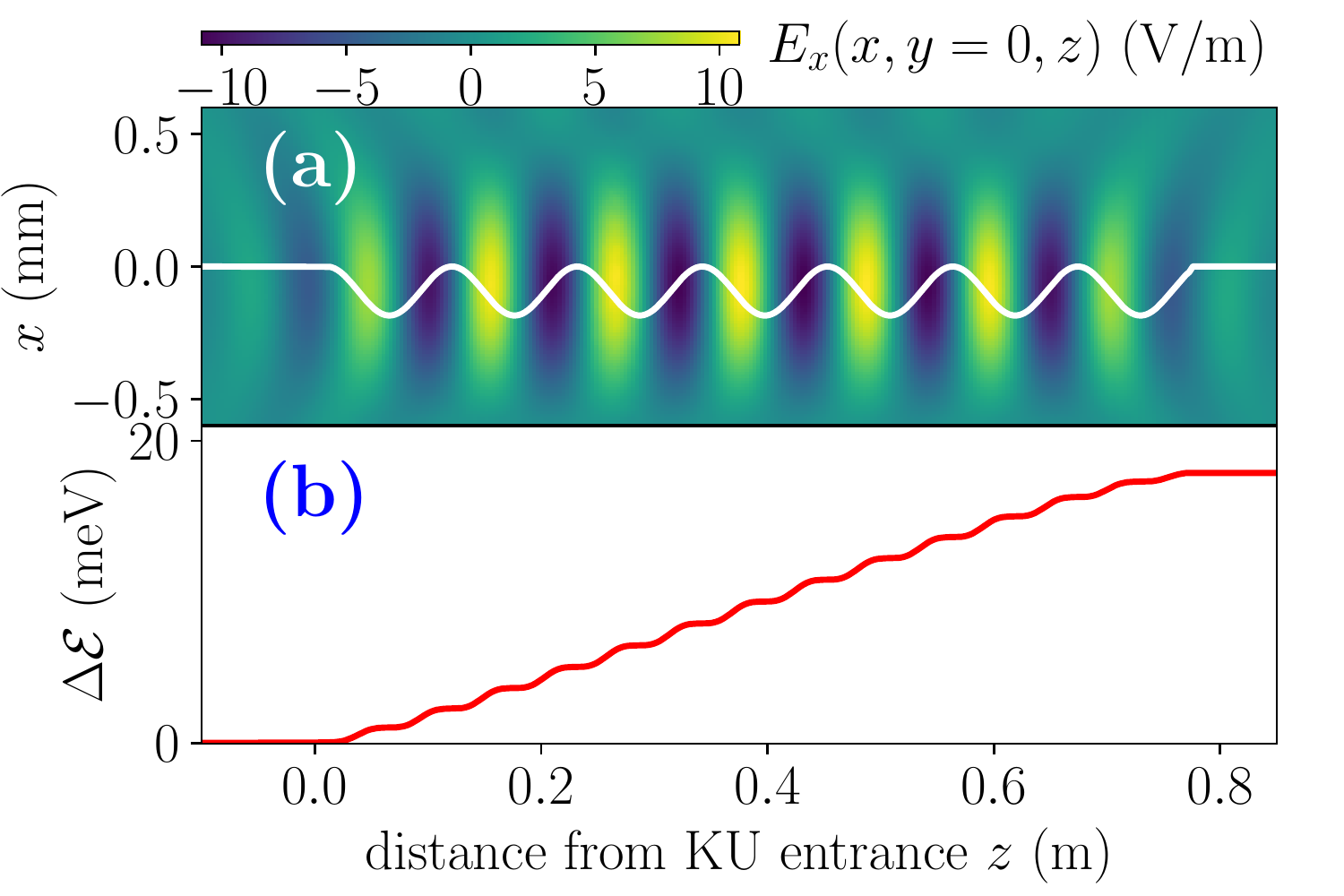}
	\caption{Transverse electric field $E_x(x,z)$, experienced by a 100-MeV electron as it moves along the KU. The white trace represents the trajectory of the electron passing through the KU (a). Corresponding evolution of the energy change along the KU for an electron phased w.r.t. the E field such to maximize its kick (b). }
	\label{fig2}
\end{figure} 
It is also being tacitly assumed that the arrival time of the particle is such that $E_x(x,z)$ maximizes the kick. A plot of the electric field in the undulator mid plane $E_x(x,y=0,z)$ appears in Fig.~\ref{fig2}(a) with the trajectory of the electron overlapped. The corresponding evolution of the electron energy along the KU is displayed in Fig.~\ref{fig2}(b). 

The kick amplitude is found to be 18~meV. A reduction of 10.4~\% comes from the longitudinal dependence of the field amplitude along the KU. The maximum transverse displacement of the particle in the KU is 93~$\mu$m allowing the particle to experience electric field values reduced by $\sim 5$~\% w.r.t. to the maximum on-axis value. Such an effect reduces the kick by only 1.1~\%. This is expected since the instantaneous energy transfer to the particle is proportional to $v_x$ which attains its maximum value on axis. As the electron's transverse offset increases the velocity decreases to eventually vanishes when the electron reaches its maximum offset. Such a dependence of the velocity $v_x(x)$ mitigates the impact of the off-axis field reduction. Furthermore for the particle receiving the largest energy kick the phase of the wave (as seen in the co-moving frame of the particle) is such that the field is zero when the particle is the farthest off axis. \\
 
\begin{figure}[h!]
 	\centering
 	\includegraphics*[width=0.48\textwidth]{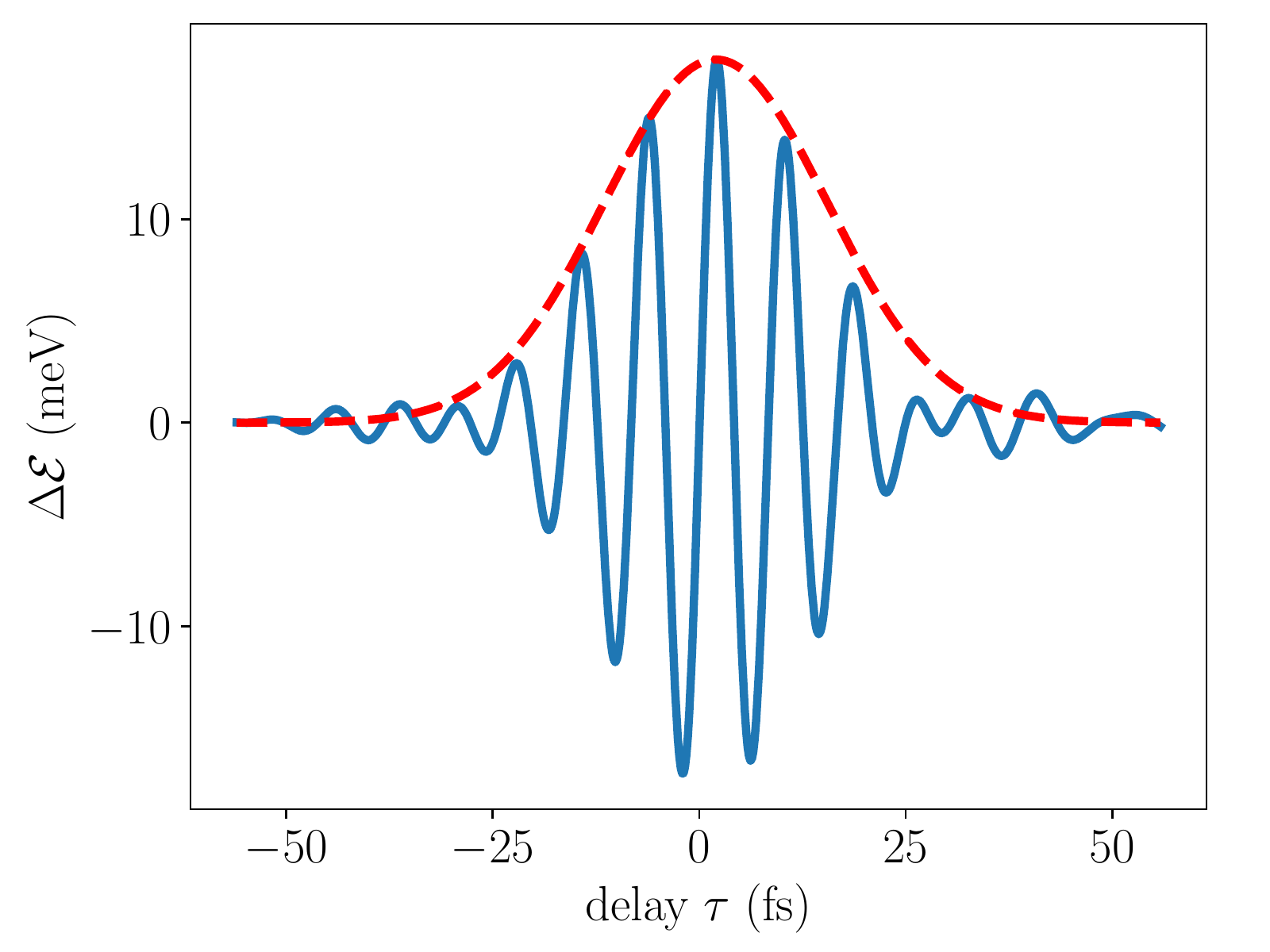}
 	\caption{Energy change as a function of the electron delay $\tau$ (solid trace). The value $\tau=0$ corresponds to the case of a reference electron which does not experience a net energy change. The dashed line is the envelope of the kick $w(\tau)$ approximated as a Gaussian function. }
 	\label{fig3}
 \end{figure}

 Our simulations also allow for the kick to be computed as a function of $\tau$ the delay relative to a reference particle as displayed in Fig.~\ref{fig3}.
The envelope, $w(\tau)$, of the kick is approximately Gaussian with a RMS duration  $\sigma_{\tau}=13.5$ fs. A common approximation to the pulse length is $t_l=N_u\lambda_0/c$ corresponding to 51.3~fs for the undulator parameters foreseen at IOTA. Since the telescope focuses light from one location in the PU to the corresponding location in the KU, the shape of the wave packet modulates while propagating through the KU. This modulation reduces the effective length of the wave packet at any particular location in the KU. Since the transverse dimensions of the wave packet ($\approx 520$~$\mu$m for the half-waist) are larger than the transverse beam size, the wave packet can be thought of as slicing the beam only along the longitudinal direction. Considering a bunch of $N$ electrons and taking the bunch density to be  constant over the length of the wave packet the number of particles within a ``sampling" slice can be approximated as
\begin{align}
N_s \simeq \frac{cN}{l_b}\int w(\tau)d\tau=\dfrac{Nc\sigma_{\tau}\sqrt{2\pi}}{l_b}
\end{align}
where $l_b$ is the bunch length. The expected bunch length in IOTA during the OSC experiment, prior to cooling, is 14.2 cm giving $N_s/N$=7.1x10$^{-5}$. In IOTA intrabeam scattering is the major limitation on the number of particles per bunch. 
\section{Conclusion}
We used a wave-optics software, SRW, to investigate the resultant energy exchange in a POSC scheme. We compared our simulation results to the semi-analytic theory developed in \cite{OSC_val} and found agreement better than $ 5 \%$ for a range of $K$ values and angular acceptances of the focusing lens. The benchmarked model was used to compute the expected kick amplitude for the POSC proof-of-principle experiment planned at the IOTA ring at Fermilab. It was especially found the decreasing of the effective aperture for points away from the kicker center reduces the energy kick by approximately $10 \%$. In addition the transverse dependence of the field experienced as the particle follows an oscillatory trajectory in the KU was found to have an insignificant effect on the energy-kick amplitude. 

So far our  calculations neglect reflective losses and dispersion from the lenses.  Accounting and compensating for dispersion is the subject of on going optimization of the optical transport for POSC in IOTA.
\section{Acknowledgments}
This work was supported by the US Department of Energy (DOE) under contract DE-SC0013761 to Northern Illinois University.  Fermilab is managed by the Fermi Research Alliance, LLC for the U.S. Department of Energy Office of Science Contract number DE-AC02-07CH11359. 
%\section*{References}
%\bibliography{mybibfile}	
	
\end{document}